\documentclass{article}
\usepackage{spconf,amsmath,graphicx}
\usepackage{bm}
\usepackage{graphicx}
\usepackage{float}
\usepackage{ascmac} 
\usepackage{fancybox}
\usepackage{float}
\usepackage{tabularx}
\usepackage{booktabs}
\usepackage{amssymb}
\usepackage{multirow}
\usepackage{cite}
\usepackage{multirow}
\usepackage[table,xcdraw]{xcolor}
\usepackage[normalem]{ulem}
\useunder{\uline}{\ul}{}
\usepackage{url}
\usepackage{stfloats}

\def\Y{{\mathbf Y}}
\def\S{{\mathbf S}}
\def\Shat{\hat{{\mathbf S}}}
\def\Stilde{\tilde{\mathbf{S}}}
\def\I{{\mathbf I}}
\def\Ihat{\hat{{\mathbf I}}}
\def\N{{\mathbf N}}
\def\Nhat{\hat{\mathbf N}}
\def\Cs{{\mathbf C}^{\text S}}
\def\Ci{{\mathbf C}^{\text I}}
\def\W{{\mathbf W}}
\def\p{\bm{p}}
\def\phat{\hat{\bm{p}}}

\def\What{\hat{\W}_{\Shat}}
\def\Wdash{\hat{\W}_{\Y}}
\def\SE{{\tt SE}}
\def\ASR{{\tt ASR}}
\def\SW{{\tt SW}}
\def\thetaSW{\theta_{\text{SW}}}
\def\SIR{\text{SIR}}
\def\SNR{\text{SNR}}
\def\CER{d}
\def\SIRe{\widehat{\text{SIR}}}
\def\SNRe{\widehat{\text{SNR}}}

\def\inRT{\in \mathbb{R}^{T}}

\def\pa{p_0}
\def\pb{p_1}

\title{Learning to Enhance or Not: Neural Network-Based Switching of Enhanced and Observed Signals for Overlapping Speech Recognition}
\name{Hiroshi Sato, Tsubasa Ochiai, Marc Delcroix, Keisuke Kinoshita, Naoyuki Kamo, Takafumi Moriya}
\address{NTT Corporation, Japan}
\begin{document}
\ninept
\setlength{\abovedisplayskip}{4pt} 
\setlength{\belowdisplayskip}{4pt}

\maketitle
\begin{abstract} 
The combination of a deep neural network (DNN) -based speech enhancement (SE) front-end and an automatic speech recognition (ASR) back-end is a widely used approach to implement overlapping speech recognition. 
However, the SE front-end generates processing artifacts that can degrade the ASR performance. 
We previously found that such performance degradation can occur even under fully overlapping conditions, depending on the signal-to-interference ratio (SIR) and signal-to-noise ratio (SNR).
To mitigate the degradation, we introduced a rule-based method to switch the ASR input between the enhanced and observed signals, which showed promising results.
However, the rule's optimality was unclear because it was heuristically designed and based only on SIR and SNR values.
In this work, we propose a DNN-based switching method that directly estimates whether ASR will perform better on the enhanced or observed signals. 
We also introduce soft-switching that computes a weighted sum of the enhanced and observed signals for ASR input, with weights given by the switching model’s output posteriors.
The proposed learning-based switching showed performance comparable to that of rule-based oracle switching.
The soft-switching further improved the ASR performance and achieved a relative character error rate reduction of up to 23 \% as compared with the conventional method.

\end{abstract}
\section{Introduction}
\label{sec:intro}
Recently, the performance of automatic speech recognition (ASR) has greatly improved, enabling it to handle more severe recording conditions including overlapping speech (speech noise) \cite{barker2018fifth} and/or background noise (non-speech noise) \cite{barker2015third}.

To cope with such conditions, various speech enhancement (SE) technologies~\cite{comon2010handbook,li2014overview,wang2018supervised} have been investigated as front-ends for ASR.
In particular, speech separation~\cite{hershey2016deep,qian2018single,luo2018tasnet} and speech extraction~\cite{wang2018deep,delcroix2018single} have been widely investigated to handle \emph{overlapping speech}. 
These techniques have led to substantial improvement in overlapping speech recognition by successfully dealing with interfering speech, even in a single-channel setup.

Despite the potential of single-channel SE technologies, DNN-based SE has a known tendency to generate ``processing artifacts'' due to non-linear transformations, which are known to be detrimental to ASR\cite{iwamoto2022how}.
For example, removal of the background noise in \emph{non-overlapping speech} regions (i.e., regions with no interfering speakers) can improve enhancement metrics such as the signal-to-distortion ratio (SDR), but it often degrades the ASR performance\cite{yoshioka2015ntt,chen2018building,fujimoto2019one}. This may be because an ASR back-end trained on multi-condition noisy data is less sensitive to noise than to processing artifacts. 
Consequently, it is a common practice to apply single-channel SE always for \emph{overlapping speech} but not for \emph{non-overlapping speech} regions~\cite{wangvoicefilter}.

However, we found in our previous work that SE degrades ASR even for \emph{overlapping speech} regions, depending on SIR and SNR in the observed mixture~\cite{sato21_interspeech}. 
From this observation, we proposed switching the ASR input between the observed and enhanced signals so as to selectively use SE only if it benefits ASR.
Through a proof-of-concept investigation, we introduced a simple rule-based switching method based on estimated SIR and SNR values.

Although the previous method improved the overall ASR performance for overlapping speech, it was certainly not optimal because the switching was implemented by a heuristic rule based only on SIR and SNR values.
In fact, as will be shown later, our detailed analysis reveals that under certain SIR and SNR conditions, SE sometimes improves ASR but sometimes does not (see Fig.~\ref{fig:distribution}). 
In other words, consideration of only the SIR and SNR values is insufficient to estimate whether SE will benefit ASR.
Accordingly, the upper-bound performance of rule-based switching is limited, even with perfect SIR and SNR estimation.
In addition, this switching can only be done in a hard-manner because it is based on thresholding.

In this work, we propose a learning-based switching method that directly estimates whether ASR will perform better on an enhanced speech or observed mixture.
Specifically, to train the DNN-based switching model, we generate training targets from overlapping speech recordings by comparing the ASR performance on the original observed mixture and the enhanced speech.
The proposed learning-based switching achieves ASR performance comparable to that of rule-based switching with oracle SIR and SNR values.

In addition, we investigate soft-switching using the continuous output of the switching model.
Because the switching model classifies whether enhancement will improve ASR, its output values can be interpreted as the posterior probabilities of each event, which include information about the uncertainty of the decision.
To incorporate this uncertainty and increase the switching robustness, the soft-switching computes a weighted sum of the enhanced speech and observed mixture with weights given by the posteriors, to generate the ASR input.
A similar formalization was also introduced for selective use of observed mixtures in \emph{non-overlapping} regions in \cite{wangvoicefilter}.
In our results, soft-switching further improves the ASR performance and achieves up to 23 \% relative reduction in the character error rate (CER) as compared with the conventional rule-based method.

\section{Related Works}
To deal with ASR performance degradation due to processing artifacts, a method called VoiceFilter-Lite with selective use of observed mixtures instead of enhanced speech was proposed~\cite{wangvoicefilter}.
A major difference between their work and ours is that they improved the ASR performance on \emph{non-overlapping regions} by scaling down the enhanced speech for those regions according to posteriors estimated by an overlap detector. In contrast, we improve ASR performance even for \emph{overlapping regions} by directly training the switching model on the basis of the ASR performance; thus, our experiments focus on fully overlapping conditions.
In future works, we plan to investigate a combination of these two approaches under partially overlapping conditions.

Other approaches to mitigate the ASR performance degradation due to SE include re-training the ASR module on enhanced speech\cite{chen2018building} and joint-training of the SE front-end and ASR back-end\cite{menne2019investigation}.
Those approaches often involve modifying the ASR back-end, which may not be possible in many practical applications. 
In contrast, our approach can be applied with black-box ASR systems that can only output recognized text from speech (e.g., speech recognition APIs).

\section{Conventional Switching Method}
In this section, we first describe the speech recognition pipeline for this study, which aims to recognize a target speaker's speech in an observed mixture. The pipeline includes a target-speech extraction module as a front-end and an ASR module as a back-end.
Then, we describe the conventional rule-based switching mechanism for overlapping speech.
\vspace{-4pt}
\subsection{Target Speech Recognition Pipeline}
In this work, we assume that the observed mixture signal $\Y \inRT$ is captured by a single microphone as $\Y = \S + \I + \N$ where $\S \inRT$, $\I \inRT$, and $\N \inRT$ denote the target signal, signal from interfering speakers, and background noise, respectively, and $T$ denotes the number of waveform samples.

The target speech extraction front-end extracts only the target signal $\S$ from the observed mixture $\Y$~\cite{delcroix2018single}. 
To indicate which speaker to extract, it is assumed that auxiliary information about the target speaker is available.
Typically, pre-recorded enrollment speech of the target speaker is used as the auxiliary information, denoted as $\Cs \in \mathbb{R}^{T_s}$ where $T_s$ is the number of samples of this information.
Given $\Cs$, the target signal $\Shat \inRT$ is estimated from the observed mixture as:
\vspace{-3pt}
\begin{align}
    \label{eq:frontend}
    \Shat = \SE(\Y, \Cs)
\end{align}
where $\SE(\cdot)$ denotes the speech extraction module.
For ASR of the target speech, the extracted signal $\Shat$ is used as the ASR input as:
\begin{align}
    \label{eq:backend}
    \What = \ASR(\Shat),
\end{align}
where $\What$ denotes the estimated transcription of the target speech, and $\ASR(\cdot)$ denotes the ASR module.

\vspace{-4pt}
\subsection{Rule-Based Switching with SIR and SNR Estimation}
Although speech extraction is an effective way to implement overlapping speech recognition, our preceding work showed that the ASR performance on observed mixtures sometimes exceeds that on enhanced speech, even under fully overlapping conditions\cite{sato21_interspeech}. 
From the observation that SIR and SNR are important cues for determining whether enhancement will improve or harm ASR, we introduced a rule-based switching method based on the estimated SIR and SNR. 
The ASR input $\Stilde \inRT$ is selected according to the following rule:
\vspace{-3pt}
\begin{align}
    \label{eq:switch}
    \Stilde =& \begin{cases}
    \Y & (\SIRe - \SNRe \geq \lambda) \\
    \Shat & (otherwise)
    \end{cases}\\
    \label{eq:sir}
    \SIRe := 10 &\log_{10} \frac{||\Shat||^{2}}{||\Ihat||^{2}} ,\;\SNRe := 10 \log_{10} \frac{||\Shat||^{2}}{||\Nhat||^{2}}
\end{align}
where $\SIRe$ and $\SNRe$ denote the estimated SIR and SNR, respectively, of the observed mixture; $\lambda$ is a threshold parameter determined on the development set; $||\cdot||$ denotes the $L^{2}$ norm and $\Ihat \inRT$ and $\Nhat \inRT$ denote the estimated interfering speech and estimated non-speech noise, respectively.
This rule reflects the observation that an observed mixture $\Y$ tends to be better for ASR than the enhanced speech $\Shat$ under high SIR and low SNR conditions.

In Eq.~\eqref{eq:sir}, $||\Ihat||^2$ and $||\Nhat||^2$ are required to estimate the SIR and SNR.
In the preceding work, target speech extraction was also applied for the interfering speaker by using the interfering speaker's enrollment speech $\Ci$, via $\Ihat = \SE(\Y, \Ci)$.
Because the extra information $\Ci$ is unavailable in some applications, the practicality of this switching is limited.
For noise, we first detected regions where both the target and interfering speakers were inactive using a voice activity detection (VAD) model. Then, we calculated the average $||\N||^2$ over the region of speaker inactivity, instead of directly calculating $||\N||^2$, assuming that the noise signal power was sufficiently stable during a speech segment. 
\vspace{-4pt}
\section{Proposed Method}
\begin{figure}[tb]
 \begin{center}
  \includegraphics[width=.95\hsize]{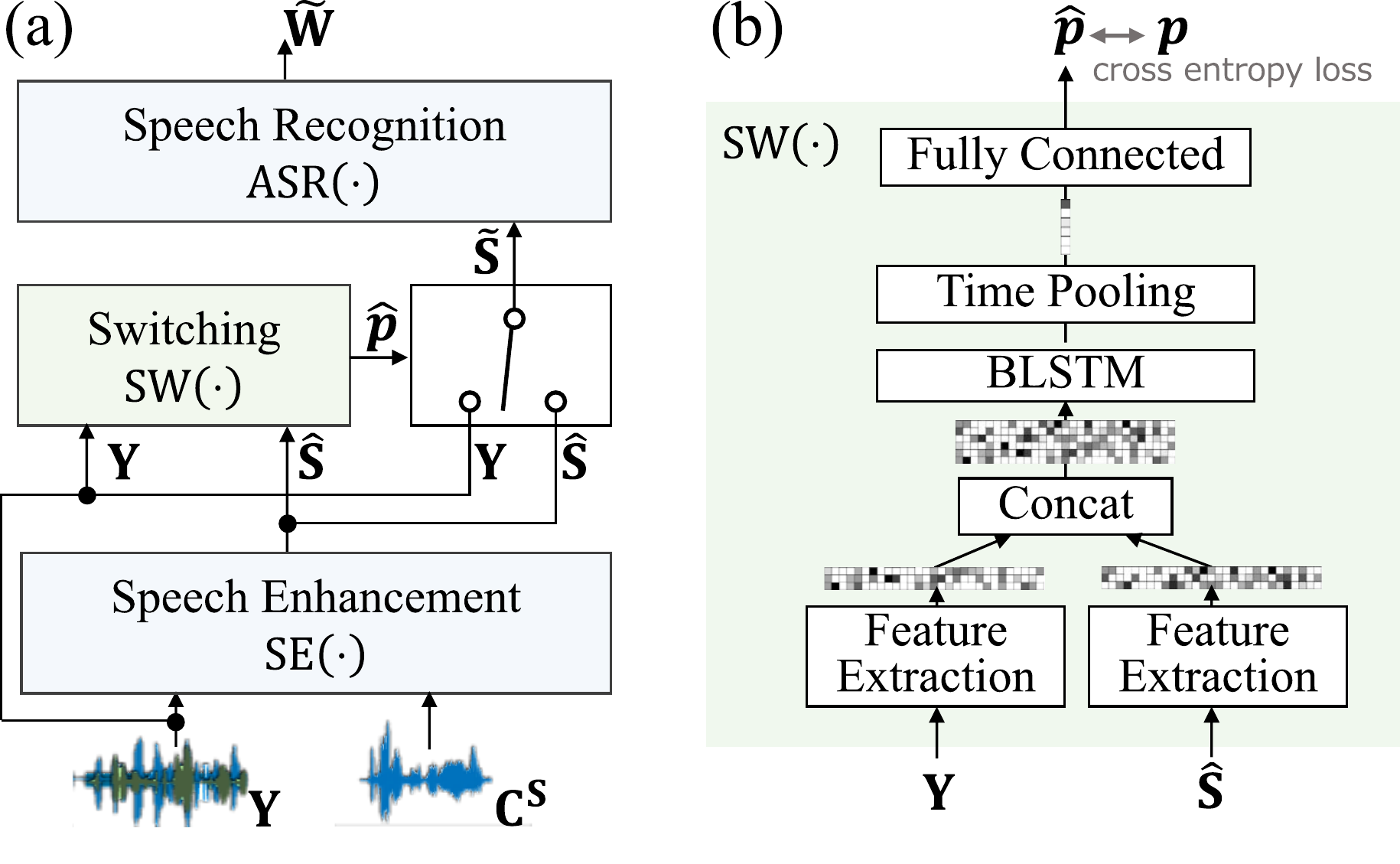}
 \end{center}
 \vspace{-24pt}
 \caption{(a) Overview of an ASR pipeline with the proposed switching model. (b) Architecture of the switching model.}
 \vspace{-12pt}
 \label{fig:framework}
\end{figure}

\begin{table*}[t]
\vspace{-6pt}
\centering
\caption{Data generation setup.}
\label{tab:setup}
\scalebox{0.91}[0.91]{
\begin{tabular}{lllcccc}
\hline
 & \multicolumn{1}{c}{dataset} & \multicolumn{1}{c}{mixture type} & \begin{tabular}[c]{@{}c@{}}SIR\\ {[}dB{]}\end{tabular} & \begin{tabular}[c]{@{}c@{}}SNR\\ {[}dB{]}\end{tabular} & \begin{tabular}[c]{@{}c@{}}\#speakers\\ (train set / dev set)\end{tabular} & \begin{tabular}[c]{@{}c@{}}\#mixtures or utterances\\ (train set / dev set)\end{tabular} \\ \hline
(a) & training data for SE & 2 speakers and noise & -5 - 5 & 0 - 20 & 777 / 49 & 50,000 / 5,000 \\
(b) & training data for ASR & 1 speaker and noise & - & 0 - 20 & 1,367 / 29 & 403,072 / 4,000 \\
(c) & evaluation data & 2 speakers and noise & 0, 10, 20 & 0, 10, 20 & 30 & 4,000 \\
(d) & training data for switching & 2 speakers and noise & -2 - 22 & -2 - 22 & 262 / 17 & 50,000 / 5,000 \\
(e) & training data for SE (for switching training) & 2 speakers and noise & -5 - 5 & 0 - 20 & 574 / 32 & 50,000 / 5,000 \\ \hline
\end{tabular}
}
\vspace{-8pt}
\end{table*}

\vspace{-4pt}
\subsection{Learning Based Switching}
The conventional rule-based switching described above showed promising results for overlapping speech recognition\cite{sato21_interspeech}.
However, the rule's optimality was unclear because it was heuristically designed and only based on SIR and SNR values.

To cope with these limitations, we propose a data-driven switching method that directly learns whether SE improves ASR or not, without explicitly estimating the SNR and SIR.
Fig.~\ref{fig:framework}(a) shows a conceptual diagram of the proposed method. 
The switching model $\SW{}(\cdot)$ performs binary classification to estimate the posterior probabilities, $\phat=[\hat{\pa}, \hat{\pb}] \in \{\hat{\pa}, \hat{\pb}: \hat{\pa}\geq0, \hat{\pb}\geq0, \hat{\pa}+\hat{\pb}=1\}$, of whether ASR will perform better on the observed mixture ($\hat{\pa}$) or the enhanced speech ($\hat{\pb}$). Given the enhanced speech and observed mixture signals, the posterior probabilities are estimated as follows:
\begin{align}
    \label{eq:posterior}
    \phat = \SW(\Y, \Shat; \thetaSW)
\end{align}
where $\thetaSW$ denotes the learnable parameters of the DNN layers in the switching model. 
From the output posteriors $\phat$, the ASR input $\Stilde$ is determined as follows:
\begin{align}
    \Stilde = \begin{cases}
    \Y & (\hat{\pa} > \hat{\pb}) \\
    \Shat & (otherwise).
    \end{cases}
\end{align}

To train the switching model, we prepare a binary classification target $\p = [\pa, \pb] \in \{[0,1],[1,0]\}$ for each instance of overlapping speech in the training dataset by measuring whether enhancement improves or degrades the ASR performance, as follows:
\begin{align}
    \label{eq:label}
    \p = \begin{cases}
    [1,0] & (\CER(\W, \Wdash) < \CER(\W, \What)) \\
    [0,1] & (otherwise)
    \end{cases}
\end{align}
where $\W$ denotes the correct transcription, $\Wdash=\ASR(\Y)$ denotes the transcription generated by recognizing the observed mixture, and $\CER(\cdot)$ denotes the edit distance between transcriptions. In this work, we adopt the CER as the distance measure.
The model parameters $\thetaSW$ are optimized by minimizing binary cross-entropy loss between the output posterior probabilities $\phat$ and the ground truth labels $\p$.

Because the proposed switching method only requires paired data of observed mixtures, their transcriptions and enrollment speech $(\Y, \W, \Cs)$, and pretrained $\ASR(\cdot)$ and $\SE(\cdot)$ modules for its training, it can potentially be optimized for real recordings in which paired clean sources and mixture recordings are unavailable and retraining of the SE module is thus difficult. 
Retraining of the switching model on such real recordings could mitigate the ASR performance degradation and enable effective application of SE to unseen real data.

\subsection{Soft Switching}
Unlike the conventional method, which deterministically switches between the enhanced speech and observed mixture by thresholding~\cite{sato21_interspeech}, the proposed learning-based switching can generate the ASR input by using the {\it soft} output posterior probabilities $\phat$. This can improve the switching robustness because it can incorporate the uncertainty in classification.
The soft-switching model is trained by the same procedure as hard-switching. For decoding, the ASR input is generated as a weighted sum of the enhanced speech and observed mixture signals, with the weights given by $\phat$ as follows:
\begin{align}
    \Stilde = \hat{\pa}\Y + (1-\hat{\pa})\Shat.
    \label{eq:soft}
\end{align}

\begin{figure}[tb]
 \begin{center}
  \includegraphics[width=1.00\hsize]{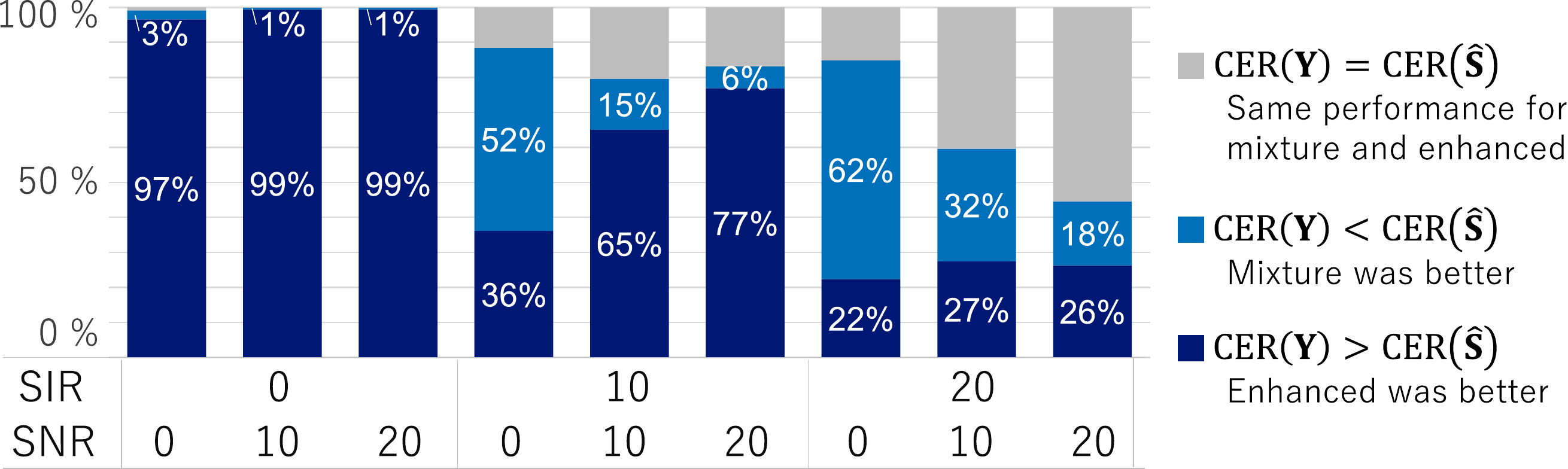}
 \end{center}
 \vspace{-20pt}
 \caption{Distribution of whether the enhanced speech or observed mixture was better for ASR, under each of the SIR and SNR conditions. The percentage on each bar indicates the ratio of utterances in which the enhanced speech or observed mixture performed better.}
 
 \label{fig:distribution}
\end{figure}

\section{Experiments}
\label{section:experiments}
\begin{table*}[t]
\centering
\caption{CER [\%] results for the observed mixture, enhanced speech, and speech selected by the conventional rule-based and proposed learning-based switching. Results are also listed for three types of oracles and for the best common soft weight $\hat{\pa}=0.9$.}
\label{tab:result}
\scalebox{0.93}[0.93]{
\begin{tabular}{cl|rrr|rrr|rrr|r}
\hline
 & \multicolumn{1}{c|}{SIR} & \multicolumn{3}{c|}{0} & \multicolumn{3}{c|}{10} & \multicolumn{3}{c|}{20} & \multicolumn{1}{c}{\multirow{2}{*}{avg.}} \\
 & \multicolumn{1}{c|}{SNR} & \multicolumn{1}{c}{0} & \multicolumn{1}{c}{10} & \multicolumn{1}{c|}{20} & \multicolumn{1}{c}{0} & \multicolumn{1}{c}{10} & \multicolumn{1}{c|}{20} & \multicolumn{1}{c}{0} & \multicolumn{1}{c}{10} & \multicolumn{1}{c|}{20} & \multicolumn{1}{c}{} \\ \hline
(1) & observed mixture & 83.9 & 80.6 & 80.7 & 20.4 & 13.8 & 16.5 & 14.8 & 6.0 & 5.3 & 35.8 \\
(2) & enhanced speech & 31.2 & 10.9 & 8.4 & 24.9 & 7.8 & 5.8 & 23.2 & 8.9 & 6.3 & 14.1 \\
(3) & soft switching with best common weight $\hat{\pa}\!=\!.9$ & 29.1 & 13.3 & 12.2 & 20.4 & 7.5 & 6.0 & 16.9 & 6.2 & 4.9 & 12.9 \\ \hline
(4) & oracle rule-based switching & 31.2 & 10.9 & 8.4 & 20.4 & 7.8 & 5.8 & 14.8 & 6.0 & 5.3 & 12.3 \\
(5) & oracle hard-switching & 30.9 & 10.8 & 8.3 & 17.4 & 6.4 & 4.9 & 13.6 & 5.1 & 4.2 & 11.3 \\
(6) & oracle soft-switching & 24.1 & 9.4 & 7.6 & 12.8 & 5.3 & 4.4 & 10.3 & 4.6 & 4.0 & 9.2 \\ \hline
(7) & rule-based switching (conventional) \cite{sato21_interspeech} & 31.2 & \textbf{10.9} & \textbf{8.4} & 21.1 & 7.8 & 5.8 & 17.0 & 7.3 & 6.1 & 12.8 \\
(8) & learnable switching (proposed) & 31.7 & \textbf{10.9} & 8.5 & 20.3 & 7.5 & 5.6 & 16.0 & 6.3 & 4.8 & 12.4 \\
(9) & \quad+ soft-switching & \textbf{30.4} & \textbf{10.9} & 8.5 & \textbf{16.5} & \textbf{6.9} & \textbf{5.5} & \textbf{13.2} & \textbf{5.7} & \textbf{4.7} & \textbf{11.4} \\
 & \quad\,\,\, (relative CER reduction from \cite{sato21_interspeech}) & (3\%) & (0\%) & (-1\%) & (22\%) & (11\%) & (5\%) & (22\%) & (21\%) & (23\%) & (11\%) \\ \hline
\end{tabular}
}
\vspace{-12pt}
\end{table*}

\subsection{Systems for Evaluation}
For experimental validation, we evaluated the ASR performance obtained by the conventional\cite{sato21_interspeech} and proposed switching methods as well as that on the enhanced speech and observed mixture, under various SIR and SNR conditions.

We also evaluated the performance achieved with three different oracle conditions: 
(i) oracle rule-based switching with perfect SIR and SNR estimation, 
(ii) oracle hard-switching, and 
(iii) oracle soft-switching.
The oracle hard-switching was obtained by choosing the signal with better ASR performance between the enhanced speech and observed mixture for each utterance.
The oracle soft-switching was obtained by determining the best soft-switching weight for each utterance. The best soft weight for $\hat{\pa}$ in Eq.~\eqref{eq:soft} was searched within the range of 0.0 to 1.0 in increments of 0.1.

To show the potential of estimating the soft switching weight for each utterance, we also evaluated the performance of soft-switching with a common weight that performed the best on average over all conditions.
The best common switching weight $\hat{\pa}$ was searched within 0.0 to 1.0 in increments of 0.1 and determined as 0.9. 

\subsection{Dataset}
All the training and evaluation were performed on simulated mixtures generated from speech recordings from the Corpus of Spontaneous Japanese (CSJ)\cite{maekawa2003corpus} and noise recordings from the CHiME-4 corpus\cite{chime4}, sampled at 16 kHz.
Table \ref{tab:setup} summarizes the data generation setup. 
Datasets (a), (b), and (c) in Table~\ref{tab:setup} were identical to datasets in our preceding work. 
The switching model was evaluated on dataset (c), given the SE model trained on (a) and the ASR model trained on (b). 

In this work, we also prepared dataset (d) to train the leaning-based switching model.
To train the switching model, we required an SE model to generate the input enhanced speech and an ASR model to generate the target labels. 
For the ASR model, we used the same model as for evaluation, which was trained on dataset (b).
On the other hand, we used an SE model trained on a different dataset than (a), which we denote as dataset (e).
The entire training data was divided to create datasets (d) for training the switching model and (e) for training the SE model, so as to train the switching model to be aware of the SE model's behaviors on unseen data.

Fig.~\ref{fig:distribution} shows the label distribution for (c) evaluation dataset at various SIR and SNR. 
Within certain SIR and SNR conditions (e.g. $(\SIR, \SNR) = (10, 0)$), SE improves ASR for some utterances but does not for others. 
Thus, it is impossible to correctly judge whether to apply SE module, based only on SIR and SNR.
To train the switching model, we omitted utterances from dataset (d) that had the same CER for the simulated mixture and enhanced speech. 
Accordingly, the actual numbers of utterances used for training and validation were 39,796 and 3,935, respectively.
We conducted experiments on almost fully overlapping dataset; the overlapping ratio of the evaluation dataset (c) was about 94 \% on average.

\subsection{System Configuration and Training Procedure}
\textbf{Switching}:
Fig.~\ref{fig:framework}(b) shows the architecture of the switching model. From the input audio signals, 256-dimensional log mel-filterbank energies were extracted on windows with a 256-sample width and a 128-sample frame shift. 
The features extracted for the enhanced speech and observed mixture were concatenated to make a 512-dimensional time series.
The features were then processed by a three-layer bidirectional long short-term memory (BLSTM) network with 128 cells, followed by a time-pooling layer and two fully connected layers with 128 cells, rectified linear unit (ReLU) activation after the hidden layer, and softmax activation after the output layer. 
We adopted attention pooling~\cite{yang2016hierarchical} for the time-pooling function.
For the conventional switching, we adopted the same threshold value of $\lambda=10$ in Eq.~\eqref{eq:switch} as in our previous work\cite{sato21_interspeech}, because the evaluation setup of this work was identical.

For optimization, we adopted the Adam optimizer\cite{kingma2014adam}. 
The initial learning rate was set as 1e-4 and was halved if the loss on the development set did not decrease for 5 epochs. The models were trained for 50 epochs, and the best-performing model based on a development set was chosen for evaluation. 

\textbf{SE front-end}:
We adopted a time-domain SpeakerBeam structure\cite{luo2019conv, delcroix2020improving} as the SE front-end module in Eq.~\eqref{eq:frontend}. Further details on the SE front-end are described in \cite{sato21_interspeech}.

\textbf{ASR back-end}:
We trained a transformer-based encoder-decoder ASR model with a connectionist temporal classification objective\cite{karita2020ctctransformer} for the ASR back-end in Eq.~\eqref{eq:backend}, according to the CSJ recipe of the ESPnet toolkit\cite{watanabe2018espnet}. The ASR system was trained on a set of unprocessed, noisy, reverberant, single-speaker utterances. Further details on the ASR back-end are also described in \cite{sato21_interspeech}.

\subsection{Results and Discussion}
Table \ref{tab:result} lists the CER results for (1) the observed mixture, (2) the enhanced speech, and speech selected by (7) the conventional and (8-9) proposed switching methods. Results for (4-6) the three types of oracles and (3) the soft-switching performance obtained by the best common switching weight $\hat{\pa}=0.9$ are also listed.

First, let us compare oracles (4) and (5) to confirm the limit of rule-based switching.
As seen in the table, there is a substantial gap in the upper-bound performance of the (4) rule-based switching and (5) learning-based hard-switching, especially under medium-to-high SIR conditions. 
This indicates that the rule-based method depending only on the SIR and SNR limited the performance of the switching.
This is most likely because the issue of whether enhancement improves or harms ASR cannot be uniquely determined for each SIR and SNR condition as revealed in Fig.~\ref{fig:distribution}. 

Second, we compare (8) the proposed method with baseline systems to show the effectiveness of the learning based approach.
Compared with (7) the conventional rule-based switching, (8) the proposed learning based method improved the CER by 0.4 \% point on average. 
Moreover, (8) our proposed method shows comparable performance to (4) the upper-bound of the conventional rule-based switching and even outperforms it under several conditions. This result indicates that the learning-based approach successfully incorporates discriminative information other than the SIR and SNR.
Besides the CER improvement, we also evaluated how well the models can classify whether speech enhancement should be used. 
The classification accuracy of (8) the proposed method was 83.0 \%, which is higher than the 80.4 \% accuracy achieved by (7) the conventional method.
As there is still a gap between the CERs achieved by (7) the proposed method and (5) the upper-bound of the hard-switching, classification improvement could further reduce the CER.

Finally, looking at (9), we can see that the proposed soft-switching achieved the best ASR performance, close to that of (5) the oracle hard-switching. The average CER was improved by 1.4 \% point from (7) the conventional switching method and by 2.7 \% point from (2) the conventional practice of always using speech enhancement for overlapping speech. 
Given the significant performance gap between (6) the oracle soft-switching and (3) soft-switching with the best common switching weight\footnote{This corresponds to performing ``observation adding'' (i.e., adding a scaled version of the observed mixture to the enhanced speech), which is a simple technique practically used to mitigate auditory distortion.}, we assume that the oracle soft weight varies for each utterance.
In contrast to the proposed soft-switching using the posterior probability of a classifier trained on hard labels, direct learning of soft-switching by using the oracle soft weight could further improve the ASR performance to be closer to that of (6) the oracle soft-switching.

\section{Conclusion}
In this work, we proposed learning-based switching between enhanced speech and observed mixtures for overlapping speech recognition. 
Our method improved the ASR performance to be comparable to that of conventional rule-based switching even with oracle SIR and SNR estimation.
We also proposed soft-switching using the posterior probability of the switching model, which achieved up to 23 \% CER reduction relative to conventional switching. 
We expect that optimization of the soft weight according to the ASR performance could further improve the performance of soft-switching.

Because the proposed switching does not require paired clean sources and mixtures for training, it can also be optimized for real data.
In our future works, we plan to apply the switching to more realistic data, including real-recorded mixtures as well as partially overlapping recordings.

\bibliographystyle{IEEEbib}
\bibliography{mybib}

\begin{thebibliography}{10}

\bibitem{barker2018fifth}
Jon Barker, Shinji Watanabe, Emmanuel Vincent, and Jan Trmal,
\newblock ``{The Fifth 'CHiME' Speech Separation and Recognition Challenge:
  Dataset, Task and Baselines},''
\newblock in {\em Proc. Interspeech}, 2018, pp. 1561--1565.

\bibitem{barker2015third}
Jon Barker, Ricard Marxer, Emmanuel Vincent, and Shinji Watanabe,
\newblock ``The third ‘chime’speech separation and recognition challenge:
  Dataset, task and baselines,''
\newblock in {\em IEEE Workshop on Automatic Speech Recognition and
  Understanding (ASRU)}, 2015, pp. 504--511.

\bibitem{comon2010handbook}
Pierre Comon and Christian Jutten,
\newblock {\em Handbook of Blind Source Separation: Independent component
  analysis and applications},
\newblock Academic press, 2010.

\bibitem{li2014overview}
Jinyu Li, Li~Deng, Yifan Gong, and Reinhold Haeb-Umbach,
\newblock ``An overview of noise-robust automatic speech recognition,''
\newblock {\em IEEE/ACM Transactions on Audio, Speech, and Language
  Processing}, vol. 22, no. 4, pp. 745--777, 2014.

\bibitem{wang2018supervised}
DeLiang Wang and Jitong Chen,
\newblock ``Supervised speech separation based on deep learning: An overview,''
\newblock {\em IEEE/ACM Transactions on Audio, Speech, and Language
  Processing}, vol. 26, no. 10, pp. 1702--1726, 2018.

\bibitem{hershey2016deep}
John~R Hershey, Zhuo Chen, Jonathan Le~Roux, and Shinji Watanabe,
\newblock ``Deep clustering: Discriminative embeddings for segmentation and
  separation,''
\newblock in {\em IEEE International Conference on Acoustics, Speech and Signal
  Processing (ICASSP)}, 2016, pp. 31--35.

\bibitem{qian2018single}
Yanmin Qian, Xuankai Chang, and Dong Yu,
\newblock ``Single-channel multi-talker speech recognition with permutation
  invariant training,''
\newblock {\em Speech Communication}, vol. 104, pp. 1--11, 2018.

\bibitem{luo2018tasnet}
Yi~Luo and Nima Mesgarani,
\newblock ``Tasnet: time-domain audio separation network for real-time,
  single-channel speech separation,''
\newblock in {\em IEEE International Conference on Acoustics, Speech and Signal
  Processing (ICASSP)}, 2018, pp. 696--700.

\bibitem{wang2018deep}
Jun Wang, Jie Chen, Dan Su, Lianwu Chen, Meng Yu, Yanmin Qian, and Dong Yu,
\newblock ``Deep extractor network for target speaker recovery from single
  channel speech mixtures,''
\newblock in {\em Proc. Interspeech}, 2018, pp. 307--311.

\bibitem{delcroix2018single}
Marc Delcroix, Kate\v{r}ina \v{Z}mol\'{i}kov\'{a}, Keisuke Kinoshita, Atsunori
  Ogawa, and Tomohiro Nakatani,
\newblock ``Single channel target speaker extraction and recognition with
  speaker beam,''
\newblock in {\em IEEE International Conference on Acoustics, Speech and Signal
  Processing (ICASSP)}, 2018, pp. 5554--5558.

\bibitem{iwamoto2022how}
Kazuma Iwamoto, Tsubasa Ochiai, Delcroix Marc, Rintaro Ikeshita, Hiroshi Sato,
  Shoko Araki, and Shigeru Katagiri,
\newblock ``How bad are artifacts?: Analyzing the impact of speech enhancement
  errors on asr,''
\newblock Submitted to ICASSP 2022.

\bibitem{yoshioka2015ntt}
Takuya Yoshioka, Nobutaka Ito, Marc Delcroix, Atsunori Ogawa, Keisuke
  Kinoshita, Masakiyo Fujimoto, Chengzhu Yu, Wojciech~J Fabian, Miquel Espi,
  Takuya Higuchi, et~al.,
\newblock ``The ntt chime-3 system: Advances in speech enhancement and
  recognition for mobile multi-microphone devices,''
\newblock in {\em IEEE Workshop on Automatic Speech Recognition and
  Understanding (ASRU)}, 2015, pp. 436--443.

\bibitem{chen2018building}
Szu-Jui Chen, Aswin~Shanmugam Subramanian, Hainan Xu, and Shinji Watanabe,
\newblock ``{Building State-of-the-art Distant Speech Recognition Using the
  CHiME-4 Challenge with a Setup of Speech Enhancement Baseline},''
\newblock in {\em Proc. Interspeech}, 2018, pp. 1571--1575.

\bibitem{fujimoto2019one}
Masakiyo Fujimoto and Hisashi Kawai,
\newblock ``One-pass single-channel noisy speech recognition using a
  combination of noisy and enhanced features.,''
\newblock in {\em Proc. Interspeech}, 2019, pp. 486--490.

\bibitem{wangvoicefilter}
Quan Wang, Ignacio~Lopez Moreno, Mert Saglam, Kevin Wilson, Alan Chiao, Renjie
  Liu, Yanzhang He, Wei Li, Jason Pelecanos, Marily Nika, and Alexander
  Gruenstein,
\newblock ``{VoiceFilter-Lite: Streaming Targeted Voice Separation for
  On-Device Speech Recognition},''
\newblock in {\em Proc. Interspeech}, 2020, pp. 2677--2681.

\bibitem{sato21_interspeech}
Hiroshi Sato, Tsubasa Ochiai, Marc Delcroix, Keisuke Kinoshita, Takafumi
  Moriya, and Naoyuki Kamo,
\newblock ``{Should We Always Separate?: Switching Between Enhanced and
  Observed Signals for Overlapping Speech Recognition},''
\newblock in {\em Proc. Interspeech}, 2021, pp. 1149--1153.

\bibitem{menne2019investigation}
Tobias Menne, Ralf Schl{\"u}ter, and Hermann Ney,
\newblock ``Investigation into joint optimization of single channel speech
  enhancement and acoustic modeling for robust asr,''
\newblock in {\em IEEE International Conference on Acoustics, Speech and Signal
  Processing (ICASSP)}, 2019, pp. 6660--6664.

\bibitem{maekawa2003corpus}
Kikuo Maekawa,
\newblock ``{Corpus of spontaneous Japanese: its design and evaluation},''
\newblock in {\em Proc. ISCA/IEEE Workshop on Spontaneous Speech Processing and
  Recognition}, 2003, p. paper MMO2.

\bibitem{chime4}
``Chime4 challenge,''
  \url{http://spandh.dcs.shef.ac.uk/chime_challenge/CHiME4}, Cited October 7
  2021.

\bibitem{yang2016hierarchical}
Zichao Yang, Diyi Yang, Chris Dyer, Xiaodong He, Alex Smola, and Eduard Hovy,
\newblock ``Hierarchical attention networks for document classification,''
\newblock in {\em Proc. of the 2016 conference of the North American chapter of
  the association for computational linguistics: human language technologies},
  2016, pp. 1480--1489.

\bibitem{kingma2014adam}
Diederik~P Kingma and Jimmy Ba,
\newblock ``Adam: A method for stochastic optimization,''
\newblock in {\em International Conference on Learning Representations (ICLR)},
  2015.

\bibitem{luo2019conv}
Yi~Luo and Nima Mesgarani,
\newblock ``Conv-tasnet: Surpassing ideal time--frequency magnitude masking for
  speech separation,''
\newblock {\em IEEE/ACM transactions on audio, speech, and language
  processing}, vol. 27, no. 8, pp. 1256--1266, 2019.

\bibitem{delcroix2020improving}
Marc Delcroix, Tsubasa Ochiai, Kate\v{r}ina \v{Z}mol\'{i}kov\'{a}, Keisuke
  Kinoshita, Naohiro Tawara, Tomohiro Nakatani, and Shoko Araki,
\newblock ``Improving speaker discrimination of target speech extraction with
  time-domain speakerbeam,''
\newblock in {\em IEEE International Conference on Acoustics, Speech and Signal
  Processing (ICASSP)}, 2020.

\bibitem{karita2020ctctransformer}
Shigeki Karita, Nelson Yalta, Shinji Watanabe, Marc Delcroix, Atsunori Ogawa,
  and Tomohiro Nakatani,
\newblock ``Improving transformer-based end-to-end speech recognition with
  connectionist temporal classification and language model integration,''
\newblock in {\em {Proc. Interspeech}}, 2019, pp. 1408--1412.

\bibitem{watanabe2018espnet}
Shinji Watanabe, Takaaki Hori, Shigeki Karita, Tomoki Hayashi, Jiro Nishitoba,
  Yuya Unno, Nelson {Enrique Yalta Soplin}, Jahn Heymann, Matthew Wiesner,
  Nanxin Chen, Adithya Renduchintala, and Tsubasa Ochiai,
\newblock ``{ESPnet}: End-to-end speech processing toolkit,''
\newblock in {\em Proc. Interspeech}, 2018, pp. 2207--2211.

\end{thebibliography}
\end{document}